\documentclass[3p,times,procedia]{elsarticle}
\usepackage{nupha_ecrc}
\usepackage{wrapfig}
\volume{00}

\firstpage{1}

\journalname{Nuclear Physics A}

\runauth{Matthew Kelsey}

\jid{nupha}

\jnltitlelogo{Nuclear Physics A}

\usepackage{amssymb}
 \usepackage{lineno}

\usepackage[figuresright]{rotating}
\usepackage{amsmath}
\usepackage{graphics}
\begin{document}

\begin{frontmatter}



%% Instructions from Editor: Please use the following \dochead only in the preprint version (e-print arXiv etc.); 
%% use empty \dochead{} when submitting to Nuclear Physics A!
\dochead{XXVIIIth International Conference on Ultrarelativistic Nucleus-Nucleus Collisions\\ (Quark Matter 2019)}
%% e.g. \dochead{17th International Conference on Dynamical Processes in Excited States of Solids}

\title{Nuclear modification factors, directed and elliptic flow of electrons from open heavy flavor decays in Au+Au collisions from STAR}


\author[label1]{Matthew Kelsey for the STAR Collaboration}
%\author[label1,label2]{Yuanjing Ji}
\address[label1]{Lawrence Berkeley National Laboratory, Berkeley, California 94720, USA}
%\address[label2]{University of Science and Technology of China, China}
\begin{abstract}
We present the analyses of single electrons from semileptonic bottom and charm hadron decays at mid-rapidity in $\sqrt{s_{NN}}$ = 200 and 54.4 GeV Au+Au collisions (talk in Heavy-Flavor session III). The data at $\sqrt{s_{NN}}$ = 200 GeV incorporate information from the Heavy Flavor Tracker which enables the topological separation of electrons originating from bottom and charm hadron decays. We report the first STAR measurements at $\sqrt{s_{NN}}$ = 200 GeV of $v_{2}$ for bottom decay electrons as a function of $p_{\text{T}}$ and $v_{1}$ for charm decay electrons as a function of electron rapidity. Additionally, we present the improved measurements of heavy-flavor decay electron $R_{AA}$ and a new measurement of the ratio of $R_{CP}$ between bottom and charm decay electrons. Finally, we also report the measurement of non-photonic electron $v_{2}$ in $\sqrt{s_{NN}}$ = 54.4 GeV data collected during the 2017 RHIC run. 
\end{abstract}


\end{frontmatter}

% \linenumbers
\section{Introduction}\label{sec:intro}

Measurements of hadrons containing Heavy-Flavor (HF) quarks are one of the key probes of the Quark Gluon Plasma (QGP) produced in heavy-ion collisions since HF quark production is restricted to the initial hard scatterings before the formation of the QGP. Parton energy loss in the QGP medium is expected to follow a hierarchy ordered by parton color charge and mass, i.e., $\Delta E(g)>\Delta E(u,d)>\Delta E(c)>\Delta E(b)$. Measurements of charm hadron nuclear modification factors ($R_{AA}$) at the RHIC ($\sqrt{s_{NN}}$ = 200 GeV) and LHC show values that are comparable to light-flavor hadrons at high transverse momentum ($p_{\rm T}$). The similar values of $R_{AA}$ can be explained by models with parton energy loss in the QGP taking into account differences in fragmentation and spectra between light and charm quarks. A systematic comparison of bottom and charm hadron $R_{AA}$ can elucidate the mass dependence of parton energy loss since models predict significantly different values for both. These proceedings present measurements of single electron $R_{AA}$ from bottom and charm semileptonic decays using the Heavy Flavor Tracker (HFT) at STAR. Additionally, decay leptons preserve the HF hadron momentum direction and are therefore excellent proxies for anisotropic flow measurements which provide information of the HF quark transport properties of the QGP, and the initial tilt of the QGP bulk in the case of HF directed flow ($v_{1})$.

\section{Bottom and Charm Decay Electron Nuclear Modification Factors}
\begin{figure}[htp]
		\centering
		       \includegraphics[scale=0.24]{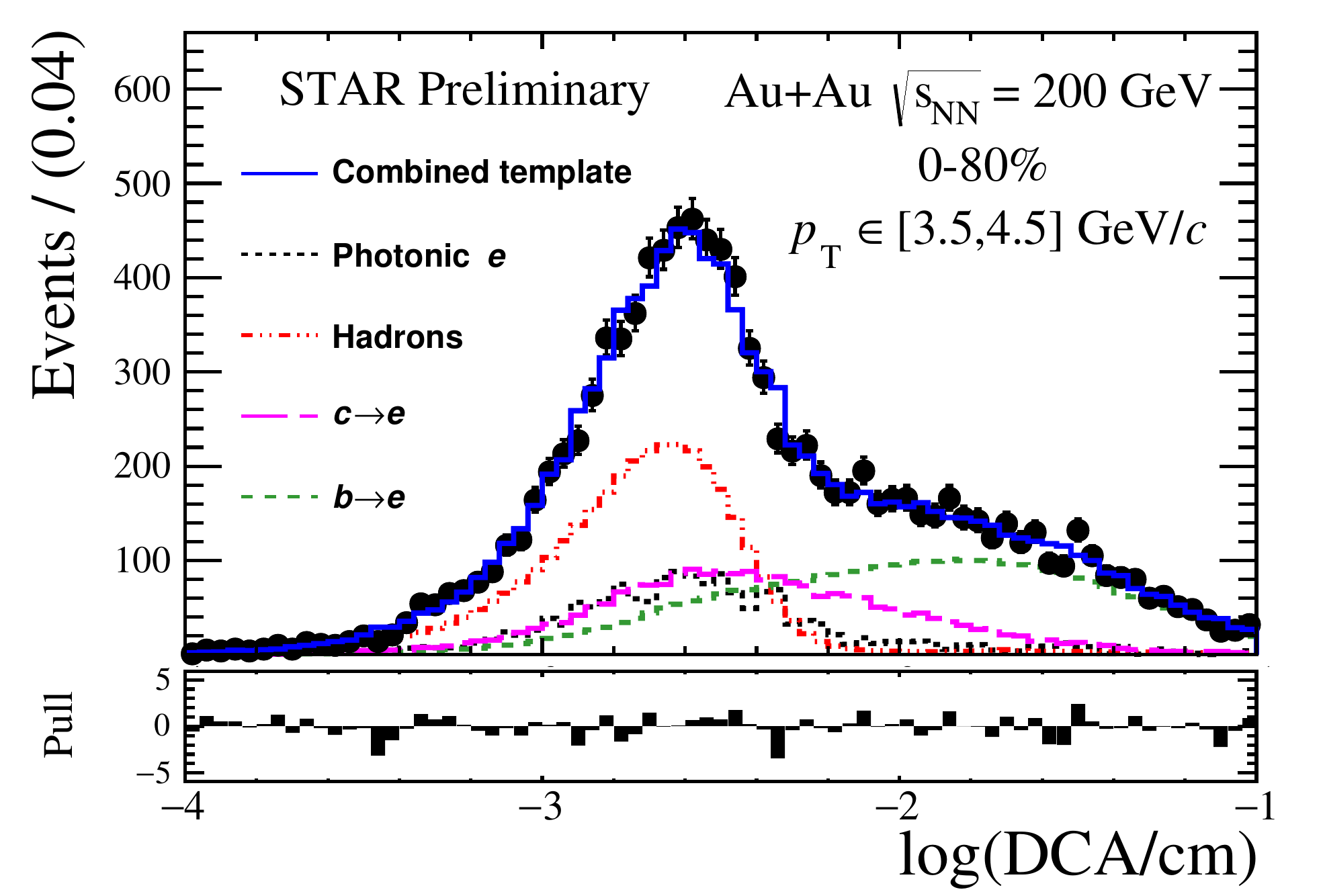}	
			\includegraphics[scale=0.24]{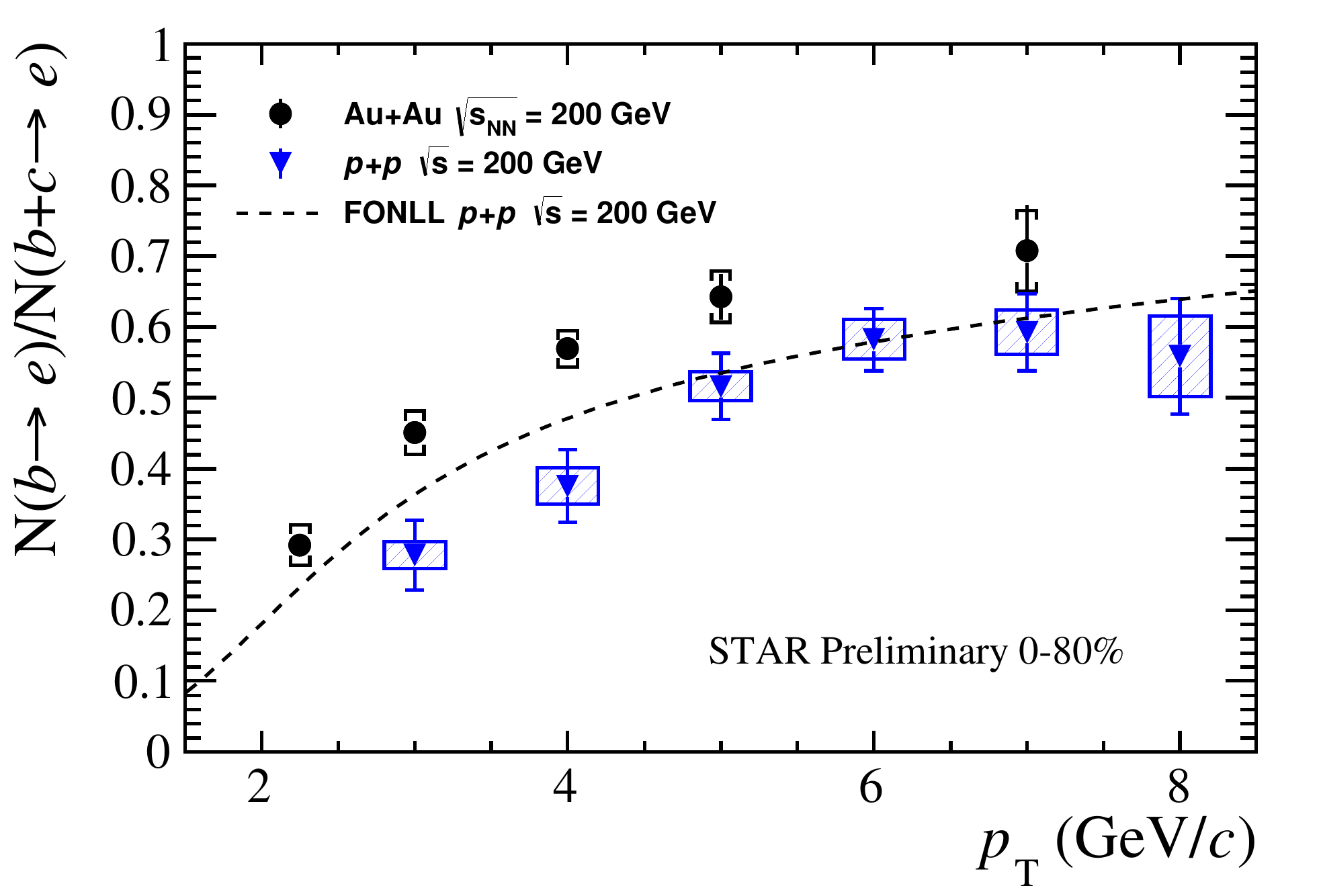}	
			\includegraphics[scale=0.24]{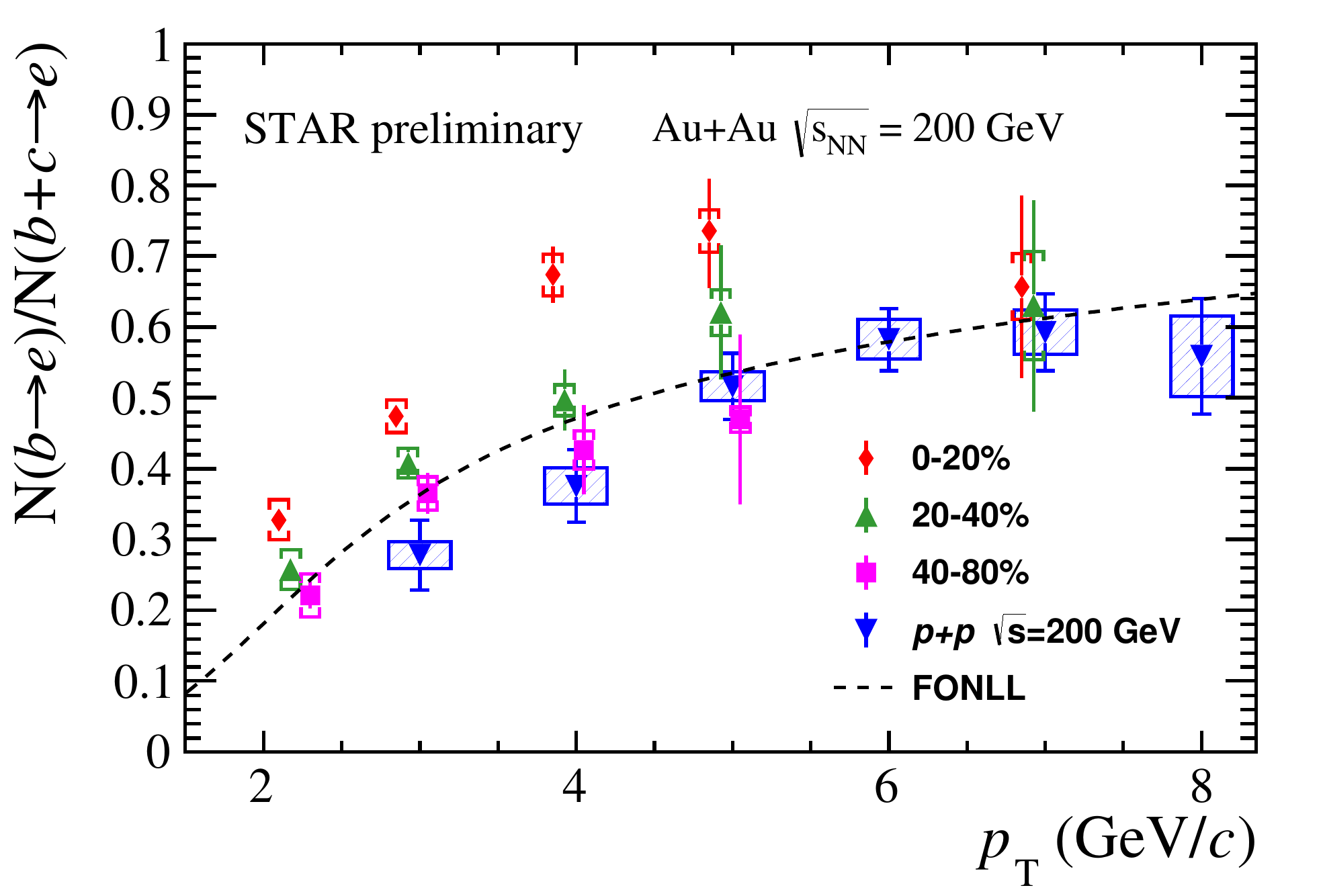}				
		\caption{Left: Example fit to the log(DCA/cm) distribution in MB $\sqrt{s_{NN}}$~=~200~GeV Au+Au collisions. Middle: The measured fraction of bottom decay electrons to all NPE electrons in MB $\sqrt{s_{NN}}$~=~200~GeV Au+Au collisions. Right: The bottom decay electron fraction as a function of $p_{T}$ in different collision centrality categories. In the middle and right panels the p+p data (\cite{PhysRevLett.105.202301} and preliminary 2012 STAR data) and the FONLL predictions~\cite{Cacciari:2012ny,Cacciari:2015fta} (black dashed line) are also shown. Statistical uncertainties are shown as error bars and systematic ones as brackets.\label{fig:bfrac}}
\end{figure}

The fraction of bottom decay electrons to the sum of all non-photonic electrons (NPE) in Au+Au collisions is extracted with a template fit to the log(DCA/cm) distribution, where the DCA is the 3D distance of closest approach of a candidate electron to the primary vertex, as shown in Fig.~\ref{fig:bfrac} left panel. The fractions of ${N(b\rightarrow e)/N(b+c\rightarrow e)}$ are shown in Fig.~\ref{fig:bfrac} middle and right panels for minimum bias (MB) and in bins of centrality, respectively, for Au+Au collisions at $\sqrt{s_{NN}}=200$ GeV. The ratios show a clear enhancement in MB and (mid-)central collisions with respect to the p+p data (\cite{PhysRevLett.105.202301} and preliminary 2012 STAR data) and the FONLL predictions~\cite{Cacciari:2012ny,Cacciari:2015fta}; The ratios in peripheral collisions are consistent with p+p data and FONLL predictions.
\begin{wrapfigure}{R}{0.5\textwidth}
%\begin{figure}[!h]
		\centering
		       \includegraphics[scale=0.38]{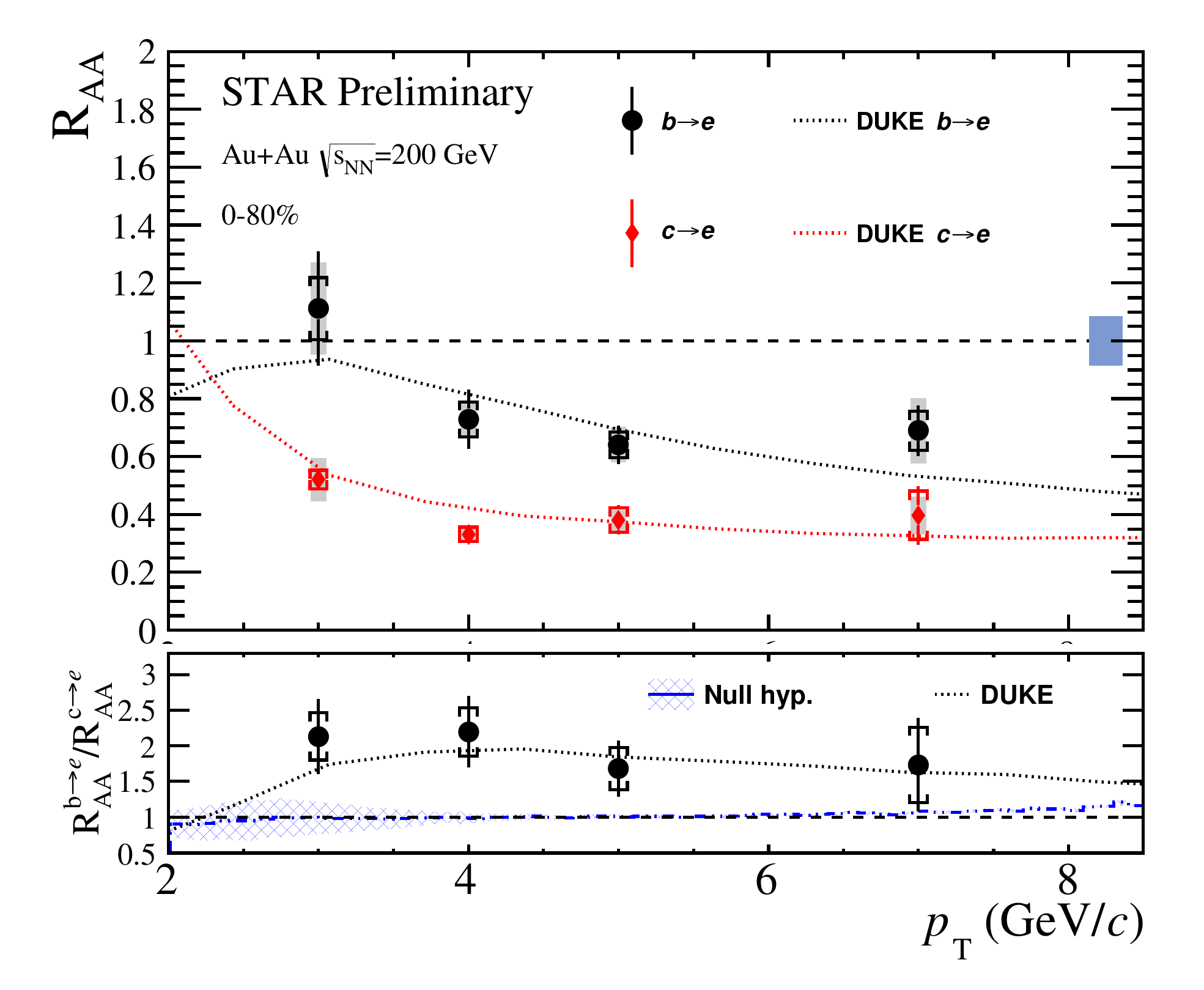}				
		\caption{The measured $R_{AA}$ for bottom and charm decay electrons as a function of electron $p_{\text{T}}$. Statistical uncertainties on the data are shown as error bars and systematic ones as brackets; the shaded gray boxes show the common uncertainty due to the inclusive NPE $R_{AA}$ measurement. The blue shaded region at $R_{AA}$=1 shows the uncertainty on N$_{coll.}$. The bottom panel shows the bottom to charm decay electron $R_{AA}$ ratio. The blue shaded curve shows the null hypothesis described in the text. In both panels Duke model~\cite{PhysRevC.92.024907} predictions are shown as the dotted lines.  \label{fig:raa}}
\end{wrapfigure}

Using the bottom decay electron fractions in Au+Au and p+p collisions, and inclusive NPE $R_{AA}$ from preliminary STAR Au+Au data, the bottom and charm decay electron $R_{AA}$ are calculated and shown in Fig.~\ref{fig:raa} along with their ratio. The data show the bottom decay electron $R_{AA}$ is larger than that of charm decay electrons, and from a constant fit to the ratio this separation is 1.92$\pm$0.25(stat.)$\pm$0.21(syst.), which is significantly different from unity at roughly a 3$\sigma$ level. A null hypothesis for the ratio (blue shaded curve) is constructed by applying D meson $R_{AA}$~\cite{PhysRevC.99.034908} to the $c/b\rightarrow e$ simulation, which takes into account the different decay kinematics in the $R_{AA}$ ratio. The p-value of the data to this curve is found to be~.014, disfavoring the hypothesis of identical charm and bottom hadron $R_{AA}$. 

We compare to the Duke Langevin transport model~\cite{PhysRevC.92.024907} shown as the dotted lines in both panels, which contains the mass dependence of energy loss and other effects which may influence the measured $R_{AA}$ (e.g., hadronization and initial HF hadron $p_{\rm T}$ spectra). Within uncertainties the model is able to describe both the absolute values of $R_{AA}$ and their ratio. This compatibility between the data and model shows a good indication that bottom quarks lose less energy in the QGP compared to charm quarks.  

An additional measurement of the ratios of bottom to charm decay electron $R_{CP}$ is performed and the data show no strong $p_{\text{T}}$ dependence. Performing the same constant fits described in the above text, we find $R_{CP}$(0-20\%/40-80\%)=1.68$\pm$0.15(stat.)$\pm$0.12(syst.) and $R_{CP}$(0-20\%/20-40\%)=1.38$\pm$0.08(stat.)$\pm$0.03(syst.). The significances of these measurements deviating from unity are 3.5$\sigma$ and 4.4$\sigma$, respectively. 
%\begin{figure}[!h]
%		\centering
%		       \includegraphics[scale=0.4]{figs/RCPDoubleRatio.pdf}	%
%			\includegraphics[scale=0.24]{figs/RCPDoubleRatioLegend.pdf}					
%		\caption{Measured data for the bottom to charm ratio of $R_{CP}$ for various centrality bins. Statistical uncertainties are shown as error bars and systematic ones as brackets. Predictions from the Duke model are shown as the respective dashed curves. \label{fig:raa1}}
%\end{figure}

\section{Bottom and Charm Decay Electron Anisotropic Flow}
\begin{figure}[htp]
		\centering
		       \includegraphics[scale=0.31]{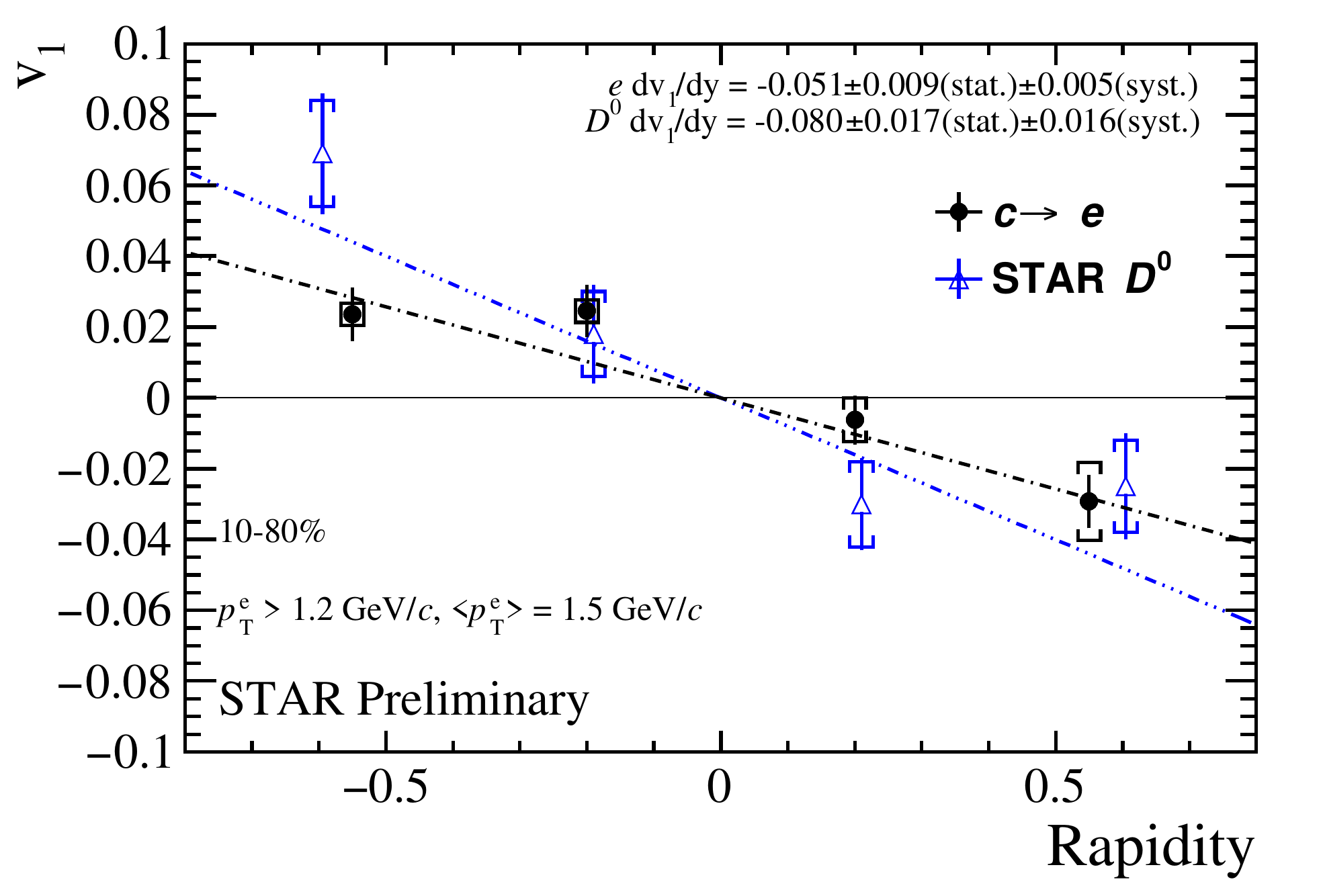}
                       \hspace{0.5cm}	
			\includegraphics[scale=0.23]{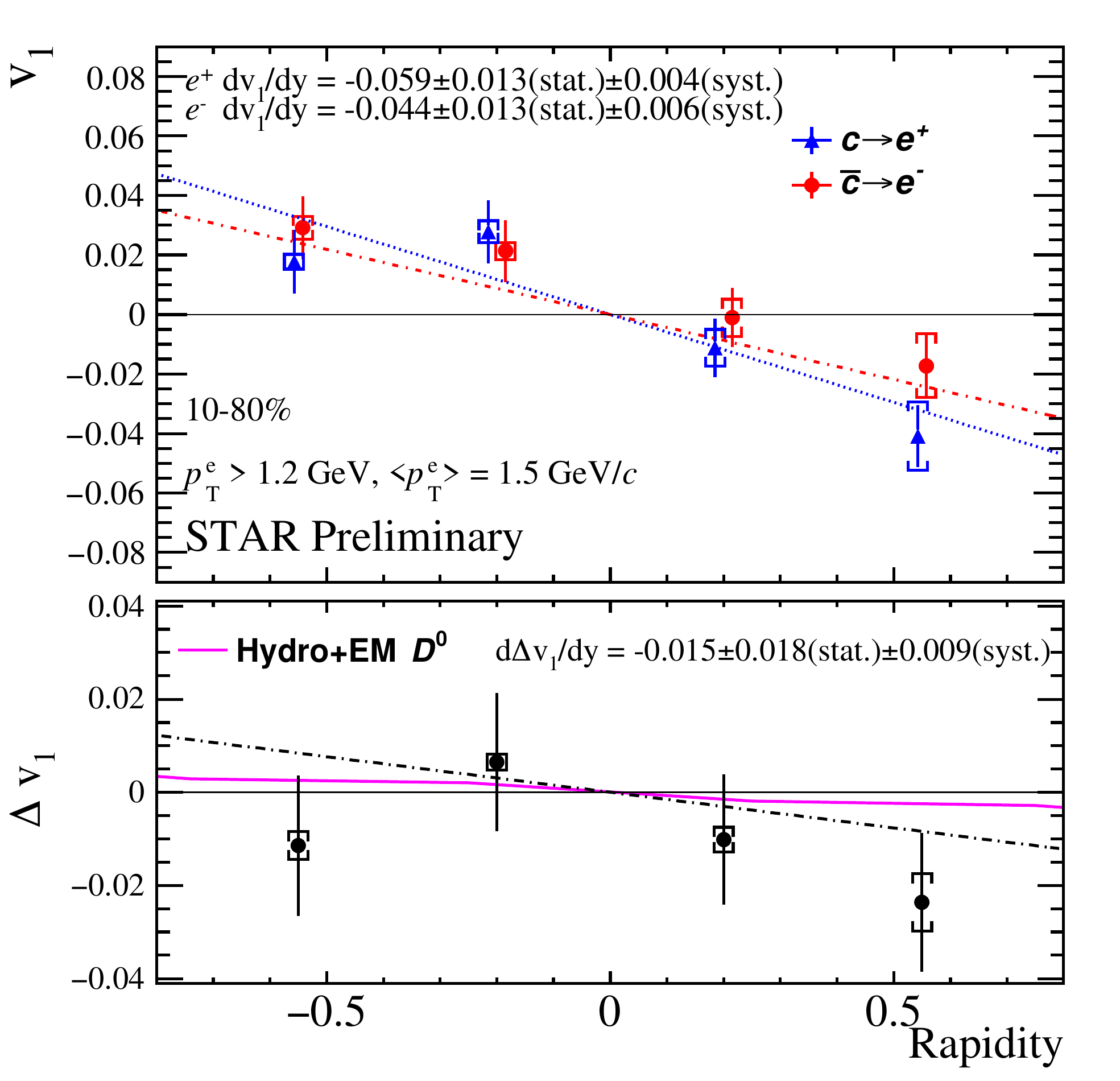}					
		\caption{Left: Measured data for the charm decay electron $v_{1}$ compared with the STAR $D^{0}$ measurement~\cite{PhysRevLett.123.162301}. Linear fits to both the electron (black dot-dash) and $D^{0}$ (blue dot-dot-dot-dash) data are shown. Right: Measured $v_{1}$ split by electron charge are shown in the top panel with linear fits to the data shown as the blue dotted and red dot-dashed lines, respectively. The bottom panel shows the difference between $e^{+}$ $v_{1}$ and $e^{-}$ $v_{1}$ and a fit to the the difference as the black dot-dashed line, and a hydrodynamic model including initial electromagnetic effects~\cite{Chatterjee_2019} as the solid magenta curve. Statistical uncertainties are shown as error bars and systematic ones as brackets.  \label{fig:v1}}
\end{figure}

The charm decay electron $v_{1}$ data are shown in Fig.~\ref{fig:v1} left and right panels for both charge averaged and split by charge, respectively. The average electron d$v_{1}$/d$y$ is in agreement with the STAR $D^{0}$ data~\cite{PhysRevLett.123.162301}, and is significant at a 5$\sigma$ level. To check for decorrelation in the $c\rightarrow e$ decay we folded the measured $D^{0}$ $v_{1}$ into simulation, and find no significant $v_{1}$ information is lost in the decay electron within the measured $p_{\rm T}$ range. The better precision of the electron measurement will be able to further constrain the charm quark drag coefficient and tilt of the QGP bulk. The difference between $e^{+}$ and $e^{-}$ $v_{1}$ is compared to a hydrodynamic model including initial electromagnetic effects~\cite{Chatterjee_2019}, and is consistent with zero within statistical uncertainties.

\begin{figure}[htp]
		\centering
		       \includegraphics[scale=0.33]{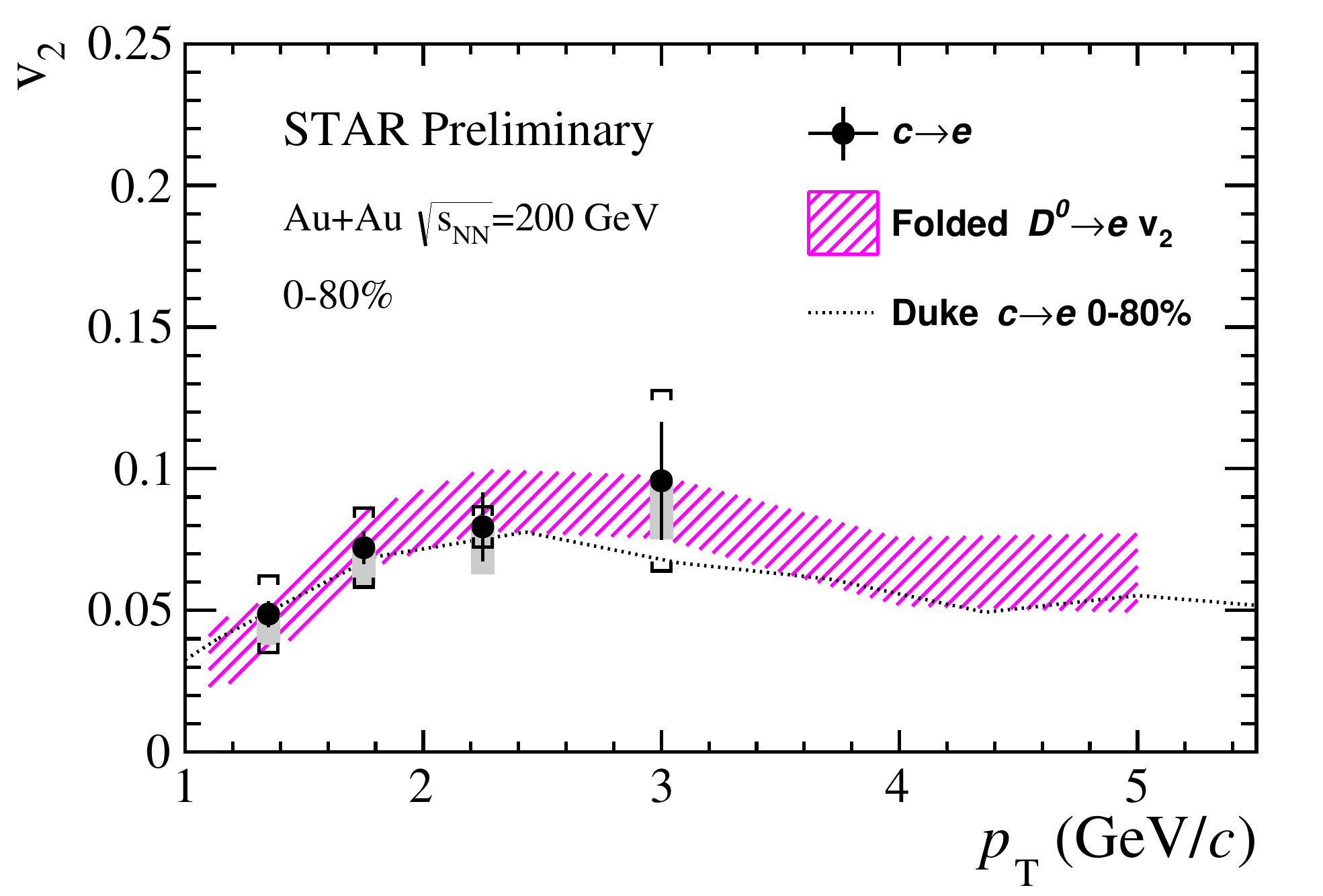}
                       \hspace{0.7cm}
		       \includegraphics[scale=0.33]{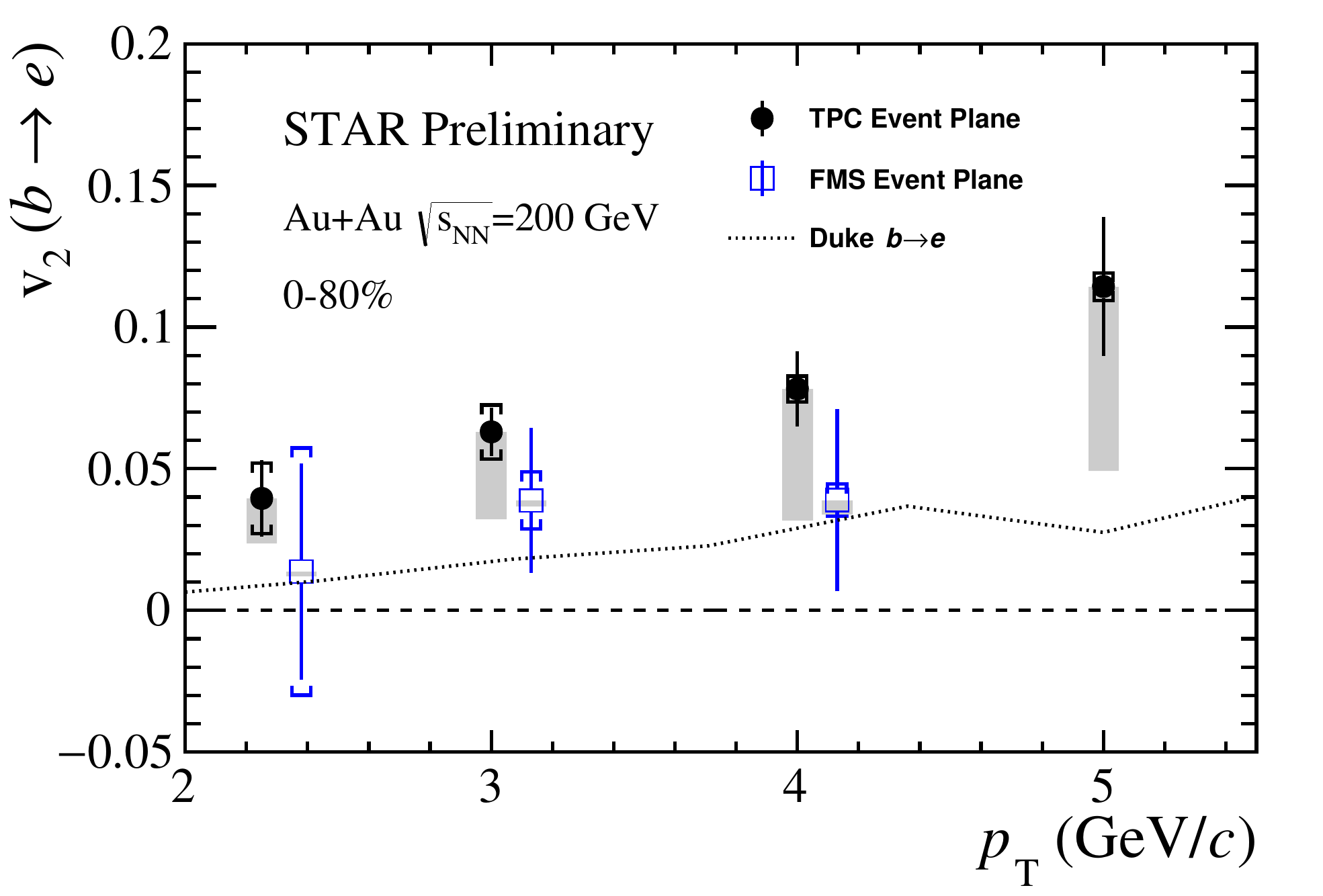}					
		\caption{Left: Measured data for the charm decay electron $v_{2}$. The folded STAR $D^{0}$ data~\cite{PhysRevLett.118.212301} are shown as the shaded magenta band (described in the text). Right: Measured bottom decay electron $v_{2}$ using the TPC (closed black circles) and FMS (open blue squares) event plane methods. Statistical uncertainties are shown as error bars and systematic ones as brackets; the gray boxes show the estimated non-flow contributions. In both panels the Duke model~\cite{PhysRevC.92.024907} is shown as the dotted black line.  \label{fig:v2}}
\end{figure}

The charm and bottom decay electron $v_{2}$ data are shown in Fig.~\ref{fig:v2} left and right panels, respectively. The contributions from non-flow effects (shown as the shaded gray boxes) are estimated using electron-hadron correlations in semileptonic charm and bottom decays in PYTHIA. The measured charm decay electron $v_{2}$ is in agreement with the measured STAR $D^{0}$ $v_{2}$~\cite{PhysRevLett.118.212301} folded to the decay electron in simulation shown as the shaded magenta band. For the bottom decay electron $v_{2}$ two event plane reconstruction methods are used with tracks reconstructed in the Time Projection Chamber (TPC, -1$<\eta<$1) or hits in the Forward Meson Spectrometer (FMS, -2.5$<\eta<$-4). In both panels the Duke model is also shown as the dotted black line, and within uncertainties is able to describe the $c\rightarrow e$ $v_{2}$ data well. After subtracting non-flow contributions, the model is also able to describe the $b\rightarrow e$ $v_{2}$ data within uncertainties. A null hypothesis ($v_{2}$=0) for the bottom decay electron $v_{2}$ using the TPC event plane with the non-flow subtracted from the central value gives a p-value of .00067 (3.4$\sigma$), indicating evidence for non-zero bottom decay electron $v_{2}$.

\section{Inclusive NPE $v_{2}$ in $\sqrt{s_{NN}}$ = 54.4 GeV Au+Au Collisions}\label{sec:54gev}

\begin{wrapfigure}{L}{0.5\textwidth}
%\begin{figure}[!h]
		\centering
		       \includegraphics[scale=0.33]{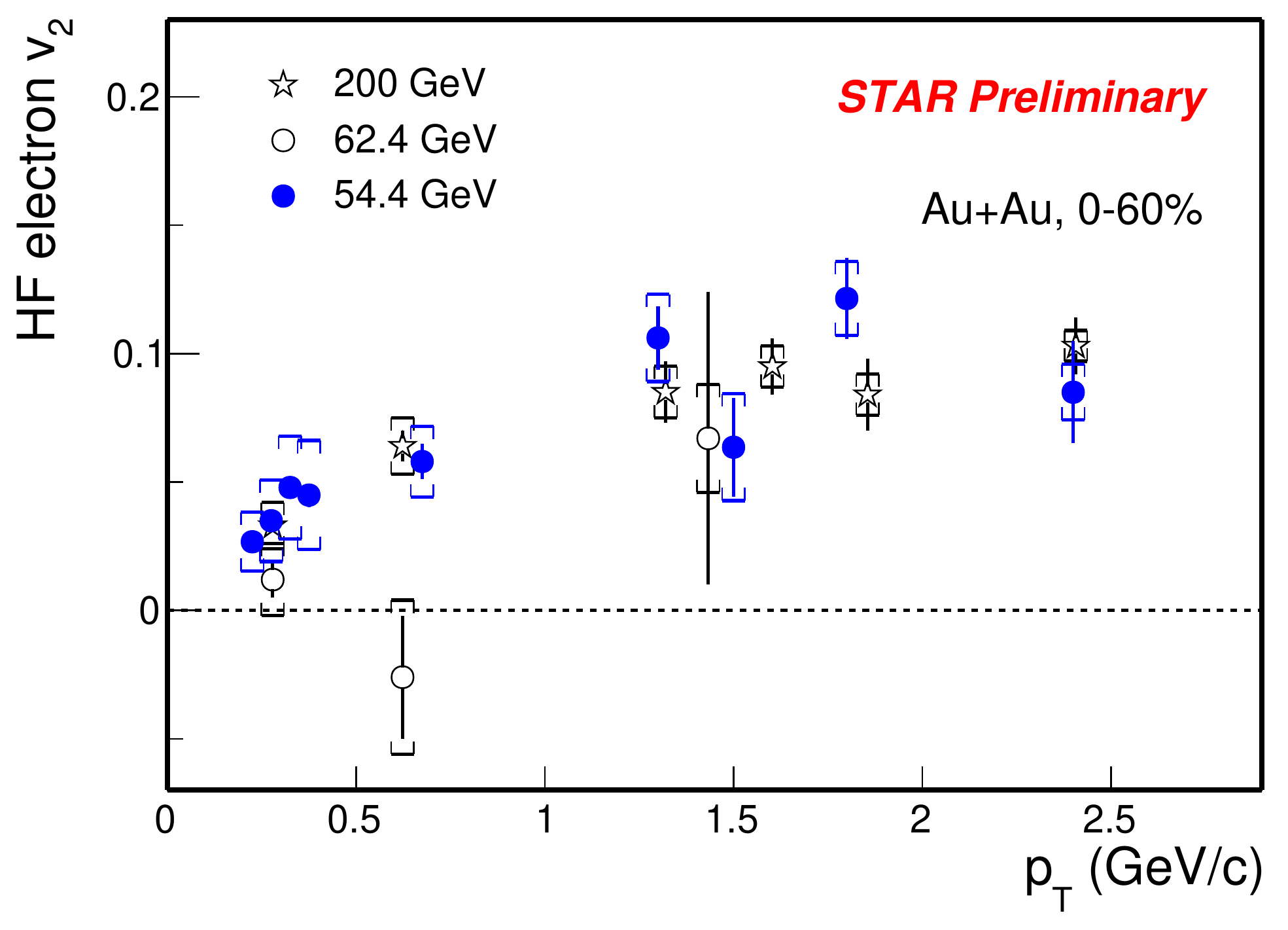}		
		\caption{Measured inclusive NPE $v_{2}$ in $\sqrt{s_{NN}}$ = 54.4 GeV Au+Au collisions. Also shown are the previous STAR measurements of inclusive NPE $v_{2}$ in $\sqrt{s_{NN}}$ = 200 and 62.4 GeV Au+Au collisions~\cite{PhysRevC.95.034907}. Statistical uncertainties are shown as error bars and systematic ones as brackets. \label{fig:v22}}
%\end{figure}
\end{wrapfigure}

The inclusive NPE $v_{2}$ is measured in $\sqrt{s_{NN}}$ = 54.4 GeV Au+Au collisions in a similar way to previous STAR measurements at $\sqrt{s_{NN}}$ = 200, 62.4, and 39 GeV~\cite{PhysRevC.95.034907}, of which the lower energy measurements were inconclusive due to the limited precision. The size of the data sample in this new measurement is roughly 15 times larger compared to the previous 62.4 GeV measurement. The data are shown in Fig.~\ref{fig:v22} with the measurements at 200 and 62.4 GeV, and the 54.4 GeV data show a significant inclusive NPE $v_{2}$ that is similar in magnitude to the data at 200 GeV. 

\section{Conclusion}\label{sec:con}

We measure the charm and bottom decay electron $R_{AA}$ using the full STAR HFT data set and find that the bottom decay electron $R_{AA}$ is larger than the charm decay electron $R_{AA}$ by roughly a factor of two with a significance of about 3$\sigma$. The measured ratios of bottom and charm decay electron $R_{CP}$ show the same hierarchy with significance greater than 3.5$\sigma$. These observations, combined with the agreement with the Duke model including mass dependence of quark energy loss, are consistent with the mass hierarchy of parton energy loss.

The data for charm decay electron $v_{1}$ and $v_{2}$ are consistent with existing measurements of $D^{0}$ by STAR and provide additional constraints on QGP transport properties, and in the former case further information on the initial tilt of the QGP.

The first measurement of significant non-zero bottom decay electron $v_{2}$ (3.4$\sigma$ from zero) at RHIC is presented. The measured $v_{2}$ values with the estimated non-flow contributions subtracted are consistent with Duke model calculations incorporating bottom quark transport in the QGP.

Finally, with the new data in $\sqrt{s_{NN}}$ = 54.4 GeV Au+Au collisions, we measure a significant inclusive NPE $v_{2}$ that is consistent with the previous STAR measurement at $\sqrt{s_{NN}}$ = 200 GeV. This measurement indicates that charm quarks interact strongly with the QGP medium in $\sqrt{s_{NN}}$ = 54.4 GeV Au+Au collisions.


%%\section{}

%% References


\bibliographystyle{elsarticle-num}
\bibliography{bib}






% \end{thebibliography}

\end{document}